\documentclass[prd,a4paper,preprint,preprintnumbers,nofootinbib]{revtex4-2}

\usepackage[a4paper, hdivide={1.91cm,,1.165cm}, vdivide={1.83cm,,3.0cm}]{geometry}

\usepackage{amsmath}
\usepackage{graphicx}
\usepackage{xspace}
\usepackage{color}
\usepackage{units}
\usepackage{slashed}
\usepackage[hyperfootnotes=false]{hyperref}
\hypersetup{linkcolor=red,
citecolor=green,
filecolor=cyan,
urlcolor=magenta}
\usepackage{gensymb}
\usepackage{booktabs}
\usepackage{multirow}
\usepackage{xcolor}
\usepackage{setspace}
\usepackage{mathtools}
\usepackage{amsfonts}
\usepackage{cleveref}

\newcommand{\ra}{\rightarrow}

\newcommand{\GG}{\langle g_s^2 G^2 \rangle}

\newcommand{\beq}{\begin{eqnarray}}
\newcommand{\eeq}{\end{eqnarray}}

\def\beq{\begin{equation}}
\def\eeq{\end{equation}}
\def\bea{\begin{eqnarray}}
\def\eea{\end{eqnarray}}

\def\vel{\left|}
\def\ver{\right|}
\def\nnb{\nonumber}
\def\lla{\left<}
\def\rra{\right>}
\def\nnb{\nonumber}
\def\la{\langle}
\def\ra{\rangle}

\def\qq{\la \bar q q \ra}

\def\es{ &=& }
\def\ar{&+& }
\def\ek{&-& }
\def\cp{&\times&}

\newcommand{\BsprimetoBstarK}{B_{s}(2S) \rightarrow B^* K}
\newcommand{\DsprimetoDstarK}{D_{s}(2S) \rightarrow D^* K}
\newcommand{\BsprimetoBsstargamma}{B_{s}(2S) \rightarrow B_s^* \gamma}
\newcommand{\DsprimetoDsstargamma}{D_{s}(2S) \rightarrow D_s^* \gamma}

\newcommand{\DsprimetoDstarKn}{D_{s}(nS) \rightarrow D^* K}
\newcommand{\BsprimetoBsstargamman}{B_{s}(nS) \rightarrow B_s^* \gamma}
\newcommand{\DsprimetoDsstargamman}{D_{s}(nS) \rightarrow D_s^* \gamma}
\newcommand{\Dfirst}{$D_{s0}^*(2317)$}
\newcommand{\Dsecond}{$D_{s1}(2460)^+$}
\newcommand{\Bfirst}{$B_{s1}(5830)^0$}
\newcommand{\Bsecond}{$B^*_{s2}(5840)^0$}

\allowdisplaybreaks

\begin{document}


\title{Strong and electromagnetic decays of the radially excited $D_s$ and $B_s$ mesons in light cone QCD sum rules}

\author{S.~Bilmis}
\email{sbilmis@metu.edu.tr}
\affiliation{Department of Physics, Middle East Technical University, Ankara, 06800, Turkey}
\affiliation{TUBITAK ULAKBIM, Ankara, 06510, Turkey}

\date{}

\begin{abstract}
  Recently, LHCb Collaboration announced the discovery of radial excitations of $D_s$ and $B_s$ mesons. In present work, we calculate the most promising strong and electromagnetic decay widths of radially excited $D_s(2S)$ and $B_s(2S)$ mesons within the light cone QCD sum rules method. 
\end{abstract}


 \maketitle
 \newpage



\section{Introduction}
\label{sec:1}
The constituent quark model has successfully described the spectroscopy of hadrons~\cite{GellMann:1964nj,Zweig:1981pd,Zweig:1964jf}. Even though most of the hadrons have already been observed, which are predicted by the quark model, many states are still waiting to be discovered. Hence, hadron-spectroscopy experiments are crucial for testing the theoretical models as well as understanding the inner structure of the hadrons. 

The anomalies observed in the experiments of charm and beauty mesons spectroscopy have received significant attention recently. After the discoveries of \Dfirst~and \Dsecond~resonances~\cite{Aubert:2003fg,Besson:2003cp} with the mass values smaller than the prediction of the potential model for cs mesons~\cite{Godfrey:2015dva}, interest in this subject increased considerably. For instance, the tetraquark~\cite{Maiani:2014aja,Browder:2003fk} and $D^* K $ molecular pictures~\cite{vanBeveren:2003kd} have been proposed. Future experimental results on the spectroscopy of cs mesons are expected to shed light on the inner structure.

Significant achievements were also obtained on the experimental side for the mesons with b-quark. The \Bfirst~ and \Bsecond~ with masses higher than the ground states $B_s^0$ and $B_s^{*0}$ mesons have been discovered~\cite{Aaltonen:2007ah,Abazov:2007af,Aaij:2012uva,Aaltonen:2013atp,Sirunyan:2018grk}. Recently, LHCb collaboration announced the observation of a new excited $D_s$ meson with mass $m = 2591 \pm 6 \pm 7~\rm{MeV}$ with quantum numbers $J^P = 0^-$ in $B^0 \rightarrow D^+ D^- K^+ \pi^-$ decay~\cite{Aaij:2020voz}. Moreover, in the analysis of the $B^+ K^-$ spectrum,  excited $B_s$ states with masses $m_1 = 6063.5 \pm 1.2 \pm 0.8$~\rm{MeV} and $m_2 = 6114 \pm 3 \pm 5$~\rm{MeV}~\cite{Aaij:2012uva} were observed.

From the theoretical point of view, these systems have huge potential in obtaining information about the non-perturbative and perturbative aspects of QCD. To determine the possible quantum numbers of these newly observed states, the measurement of the mass is not enough. Strong and electromagnetic decays of these mesons play a crucial role in understanding the structure of these mesons and establishing the possible quantum numbers. Hence, to identify the structure of these mesons, the strong coupling constants of $\DsprimetoDstarK$, $\BsprimetoBstarK$ decays as well as the decay constants of radiative $\DsprimetoDsstargamma$ and $\BsprimetoBsstargamma$ decays within light cone sum rules (LCSR) are calculated by assuming that $D_s(2S)$ and $B_s(2S)$ are the first radial excitation of $D_s$ and $B_s$ mesons. Using the obtained results for these coupling constants, we estimated the decay widths of the corresponding transitions. 

The paper is organized as follows. In section~\ref{sec:2}, the light cone sum rules for the relevant decay constants of the corresponding transitions are derived. The numerical analysis of the obtained results for the coupling constants is presented in section~\ref{sec:3}. Moreover, we also presented the values of the corresponding decay widths in this section. The final section contains our conclusion.

\section{The light cone sum rules for the electromagnetic and strong decays of $D_{s}(2S)$ and $B_{s}(2S)$ mesons}
\label{sec:2}

First, let us focus on the calculation of the strong coupling constant for $D_{s}(nS) (B_{s}(nS)) \rightarrow D^* (B^*) K$ transitions where $n=1(2)$ corresponds to the ground (first radial excited) state. The $D_{s}(nS) D^* K$ coupling is defined by the matrix element
\begin{equation}
  \label{eq:1}
  \langle D^*(p) K(q) | D_{s}(nS) (p+q) \rangle = g_n(q \epsilon)~,
\end{equation}
where $\epsilon$ is the polarization vector of the $D^*$ meson and momentum of the corresponding particles is presented in brackets. The corresponding coupling for $B_s(nS)$ is obtained by replacing $c \rightarrow b$, $D^* \rightarrow B^*$ and $D_{s}(nS) \rightarrow B_{s}(nS)$. Note that, even though $D_{s}(1S) \rightarrow D^* K$ is kinematically forbidden, this interaction still contributes to the sum rule, hence, needs to be taken into account for the calculation of the amplitude.

The coupling constant in the framework of QCD sum rules is obtained by matching the representations of the corresponding correlation function in terms of the hadrons and quark-gluons. For this purpose, we consider the following correlation function
\begin{equation}
  \label{eq:2}
  \Pi_{\mu}(p,q) = i \int d^4 x e^{ipx} \langle K(q) | T \{ \bar{d} \gamma_\mu c(x) \bar{c}(0) i \gamma_5 s(0) \} | 0 \rangle .
\end{equation}

The representation of the correlation function in terms of the hadrons is obtained by inserting the complete set of hadrons carrying the same quantum numbers as the interpolating current and isolating the contributions of $D_{s}(nS)$ and $D^*$ states. Hence, we get the following expression for the correlation function from hadronic side
\begin{equation}
  \label{eq:3}
  \Pi_{\mu}(p,q) = \sum_{n=1}^{2} \frac{ \langle 0 | \bar{q} \gamma_\mu c(x) | D^*(p) \rangle \langle D^*(p) K(q) | D_{s}(nS)(p+q) \rangle \langle D_{s}(nS) | \bar{c}(0) i \gamma_5 s(0) | 0 \rangle}{({m^2_{D^*}} - p^2) \big(m_{D_{s}(nS)}^2 - (p+q)^2\big)}~.
\end{equation}

The matrix elements entering to Eq.~\eqref{eq:3} are defined as
\begin{equation}
  \label{eq:4}
  \begin{split}
    \langle 0 | \bar{d}(x) \gamma_\mu c(x) | D^{*}(p) \rangle &= f_{D^*} m_{D^*} \epsilon_\mu~, \\
    \langle D_{s}(nS) | \bar{c}(0) i \gamma_5 s(0) | 0 \rangle &= \frac{m^2_{D_{s}(nS)} f_{D_{s}(nS)}}{m_c + m_s}~. 
  \end{split}
\end{equation}
Inserting Eq.~\eqref{eq:1} and Eq.~\eqref{eq:4} into Eq.~\eqref{eq:3} and performing summation of ${D}^{*}$ meson polarizations, we get the following expression from the hadronic part
\begin{equation}
  \label{eq:5}
  \begin{split}
    \Pi_\mu (p,q) &= \sum_{n=1}^{2} \frac{f_{{D}^*} m_{{D}^*} f_{D_{s}(nS)} m^2_{D_{s}(nS)}}{m_c + m_s} \frac{g_n}{\big(m_{D^*}^2 - p^2\big) (m_{D_{s}(nS)}^2 - (p+q)^2)} \\
    & \times \big\{
      q_\mu - \frac{p_\mu}{2 m_{D^*}^2} ( m_{D_{s}(nS)}^2 - m_{D^*}^2 - m_K^2) \big \}~. 
  \end{split}
\end{equation}
Note that, to determine the strong coupling constants, $g_n$, for $D_s(nS) \rightarrow D^* K$ transition, we choose the structure $q_\mu$. The calculation of the correlation function in terms of the quark-gluon degrees of freedom is calculated in deep Euclidean region where both virtualities  $p^2$ and $(p+q)^2$ are negative and large, hence $c$-quark is far off-shell.

After applying Wick theorem for the theoretical part of the correlation function, we get
\begin{equation}
  \label{eq:6}
  \Pi_{\mu}(p,q) = i \int d^4x \langle K(q) | \bar{d}^a(x) S^{ab}(x) i \gamma_5 s^b(0) | 0 \rangle~.
\end{equation}

In the presence of external background field, the heavy quark propagator in x-representation is given as
\begin{equation}
  \label{eq:7}
  \begin{split}
    S_{\alpha \beta}^{a a^\prime} (x) &= \frac{m_Q^2}{4 \pi^2} \bigg\{ \frac{K_1 (m_Q \sqrt{-x^2})}{(\sqrt{-x^2})^2} + \frac{i \slashed{x} K_2 (m_Q \sqrt{-x^2})}{(\sqrt{-x^2})^2} \bigg\}_{\alpha \beta} \delta^{a a^\prime} \\
    & - \frac{g_s m_Q}{16 \pi^2} \int_0^1 du \bigg[ \frac{iK_1 (m_Q \sqrt{-x^2})}{\sqrt{-x^2}} (u \slashed{x} \sigma_{\alpha \beta} + \bar{u} \sigma_{\alpha \beta} \slashed{x}) + K_0 (m_Q \sqrt{-x^2}) \sigma_{\lambda \tau} \bigg]_{\alpha \beta} {G^{(l)}}^{\lambda \tau} \big(\frac{\lambda^l}{2}\big)^{a a^\prime}
\end{split}
\end{equation}
where $G_{\lambda \tau}^{(l)}$ is the gluon field strength tensor, the $\lambda^l$ are the Gell-Mann matrices and $K_i(m_Q\sqrt{-x^2})$ are the modified Bessel functions of the second kind.

From Eq.\eqref{eq:6}, it follows that the calculation of the theoretical part of the correlation function reduces to the determination of the matrix elements $\langle K(q) | \bar{d} \Gamma_i s | 0 \rangle $ and $\langle K(q) | \bar{d} \Gamma_i G_{\lambda \tau}^{(l)} s | 0 \rangle $ after using the following Fierz identities
\begin{equation}
  \label{eq:8}
  \begin{split}
    q_\alpha^{a} \bar{q}_\beta^{a^\prime} &= - \frac{1}{12}(\Gamma_i)_{\alpha \beta} \delta^{a a^\prime} \bar{q} \Gamma_i q~, \\
    q_\alpha^{b} \bar{q}_\beta^{b^\prime} G_{\lambda \tau}^{(l)} &= - \frac{1}{16}\big(\frac{\lambda^{(l)}}{2}\big)^{b b^\prime}  (\Gamma_i)_{\alpha \beta} \bar{q} (\Gamma_i) q G_{\lambda \tau}^{(l)}~,  
\end{split}
\end{equation}
in which $\Gamma_i$ is the full set of Dirac matrices, $\Gamma_i = \{ I, \gamma_5, \gamma_\mu, i \gamma_\mu \gamma_5, {\sigma_{\mu \nu} \over \sqrt{2}} \} $. 
These matrix elements are the main non-perturbative  ingredients of the light cone sum rules. The matrix elements $\langle K(q) | \bar{q} \Gamma_i s | 0 \rangle$ and $\langle K(q) | \bar{q} \Gamma_i G_{\lambda \tau}^{(l)} s | 0 \rangle$   are parameterized in terms of K-meson distribution amplitudes(DA) of different twists. These expressions are given in~\cite{PhysRevD.51.6177,Khodjamirian:2020mlb,Ball:2006wn,Ball:2004ye,Ball:1998sk} and we present these DA's in Appendix A for completeness.

Inserting Eqs.\eqref{eq:7} and \eqref{eq:8} into Eq.\eqref{eq:6}, and performing Fourier transformation first, and double Borel transformation over $-p^2$ and $-(p+q)^2$ after in both representations of the correlating function and matching the coefficients of the structure $q_\mu$ for $\DsprimetoDstarKn$ coupling constant, we get the following sum rule
\begin{equation}
  \label{eq:9}
  \begin{split}
  \Pi^{theor.} =  & g_1 \frac{f_{D^*} m_{D^*} f_{D_{s}(1S)} m_{D_{s}(1S)}^2}{m_c+m_s}  e^{- (m_{D_{s}(1S)}^2 / {M_1^2} + m_{D^*}^2 / {M_2^2}) } + \\
      & g_2 \frac{f_{D^*} m_{D^*} f_{D_{s}(2S)} m_{D_{s}(2S)}^2}{m_c + m_s} e^{- (m_{D_{s}(2S)}^2 / {M_1^2} + m_{D^*}^2 / {M_2^2}) }~,
\end{split}
\end{equation}
where $\Pi^{theor.}$ is presented in Appendix B and $M_1^2$, $M_2^2$ represent the Borel mass parameters for the initial and final state channels, respectively.

Since the masses of the initial and final state mesons are nearly the same, we can use $M_1^2 = M_2^2 = 2 M^2$. In result, we have one equation but two unknowns, $g_1$ and $g_2$. To obtain the second equation, we get the derivative of both sides of the equation~\eqref{eq:9} with respect to $-1/M^2$. In the following discussions, since we focus on the strong decay of the radially excited meson to the $D_{s}(2S) \rightarrow D^* K$, we just deal with the coupling constant $g_2$. By solving the two equations, we obtain $g_2$ as
\begin{equation}
  \begin{split}
    g_2 & = - \frac{(m_c + m_s) e^{\frac{m_{D_{s}(2S)}^2 + m^2_{D^*}}{2 M^2}} \big( (m_{D_{s_1}^2}  + m^2_{D^*}) \Pi - 2 \Pi^\prime \big)}{f_{D_{s}(2S)} f_{D^*} m_{D_{s}(2S)}^2 m_{D^*} (m_{D_{s}(2S)}^2 - m_{D_{s}(1S)}^2) }~,
  \end{split}
\end{equation}
where $\Pi^\prime$ denotes the derivation with respect to $-\frac{1}{M^2}$ of $\Pi$.

The results obtained here can be improved by taking $\mathcal{O}(\alpha_s)$ corrections. Moreover, once we use $m_{K}^{2} \rightarrow 0$ and by setting $g_2 = 0$ and $m_s \rightarrow 0$ from our results, we can obtain the results for $g_{D^* D \pi}$ coupling which are calculated with and without $\mathcal{O}(\alpha_s)$ contributions in~\cite{PhysRevD.51.6177} and \cite{Khodjamirian:2020mlb}, respectively.

Now, let us turn our attention to the calculation of the coupling constant $f_n$ for the $\DsprimetoDsstargamman$ ($\BsprimetoBsstargamman$) transition. The transition matrix element between $D_{s}(nS)$ and $D_s^*$ states due to the electromagnetic current is defined as
\begin{equation}
  \label{eq:16}
  \langle D_s^*(p)| j_{el}^\mu | D_{s}(nS) (p+q) \rangle = \epsilon^{\mu \nu \alpha \beta} p_\nu \epsilon_\alpha q_\beta f_n(q^2)~,
\end{equation}
where $f_n$ is the transition amplitude and $\epsilon$ is the four-vector polarization of the $D_s^*$ meson. Since the emitted photon is real in this decay, then we need the value of $f_n(q^2)$ only at $q^2 = 0$ point. The
$D_s(nS) (B_s(nS)) \rightarrow D_s^* (B_s^*)\gamma$  decay is described by the following correlation function
\begin{equation}
  \label{eq:17}
  \Pi^{\mu \nu} (p,q) = i^2 \int d^4 x d^4 y e^{i p x + i q y} \langle 0 | T \{j^{\mu}_{D_s^* (B_s^*)} (x) j_{el}^\nu(y) j_{D_{s}(nS)(B_{s}(nS))}(0)\} | 0 \rangle~,
\end{equation}
where $j_{el}^{\nu} = e_q \bar{q} \gamma^{\nu} q + e_Q \bar{Q} \gamma^\nu Q$ and $q (Q)$ is the light (heavy) quarks and $e_q (e_Q)$ denotes its charge. The interpolating currents of $D_{s}^* (B_{s}^*)$ and $D_{s}(nS) (B_{s}(nS))$ mesons are 
\begin{equation}
  \label{eq:18}
  \begin{split}
  j^\mu_{D_s^* (B_s^*)} &= \bar{s}^{a}(x) \gamma^\mu Q^a(x) \\
  j_{D_{s}(nS) (B_s(nS))} &= \bar{s}^{b}(x) i \gamma^5 Q^b(x) 
\end{split}
\end{equation}
where $Q^a = c(b)$ for $D_s(B_s)$ case and $a$ and $b$ are the color indices. Hence, we obtain the phenomenological part of the correlation function as
\begin{equation}
  \label{eq:19}
  \Pi^{\mu \nu} = \sum_{n=1}^{2 }f_n \frac{f_{D_s^*} f_{D_{s}(nS)} m_{D_s^*}}{m_c + m_s}  \frac{m_{D_{s}(nS)}^2}{(p^2 - m_{D_s^*}^2) ((p+q)^2 - m_{D_{s}(nS)}^2)} \epsilon^{\mu \nu \rho \beta}p_\rho q_\beta + ...
\end{equation}
Here dots describe the contributions of excited states and continuum. To derive the Eq.~\eqref{eq:19}, we used the standard definitions given by Eq.~\eqref{eq:4}.

By introducing the electromagnetic background, the correlation function can be rewritten in the following form
\begin{equation}
  \label{eq:21}
  \Pi^{\mu \nu} \eta_\nu = i \int d^4 x e^{i p x} \langle 0 | T \{ j^{\mu}_{D_s^*}(x) j_{D_{s}(nS)}(0) \} |0 \rangle_F~,
\end{equation}
where subscript $F$ denotes that the vacuum expectation values are evaluated in the presence of the background field
\begin{equation}
  \label{eq:22}
  F_{\mu \nu} = i (\eta_\mu q_\nu - \eta_\nu q_\mu) e^{i q x}~,
\end{equation}
in which $\eta_\mu$ is the photon polarization four vector.

On the other hand, at hadronic level, the expression for the correlation function can be obtained from Eq.~\eqref{eq:19} by multiplying it with $\eta_\nu$
\begin{equation}
  \label{eq:23}
  \Pi^{\mu \nu} \eta_\nu = \sum_{n=1}^{2} f_n \frac{f_{D_s^*} f_{D_{s}(nS)} m_{D_s^*} m_{D_{s}(nS)}^2}{m_c + m_s} \frac{1}{(p^2 - m_{D_s^*}^2) ((p+q)^2 - m_{D_s^*}^2)} \epsilon^{\mu \nu \rho \beta} \eta_\nu p_\rho q_\beta~.
\end{equation}
As we noted earlier, to construct the sum rules for the relevant quantity, calculation of the correlation function in the deep Euclidean domain is needed. This can be performed by inserting the explicit expressions of the interpolating currents into the Eq.~\eqref{eq:21}

\begin{equation}
  \label{eq:24}
  \Pi^{\mu \nu}(p,q) \eta_\nu = i \int d^4x e^{i p x} \langle 0 | T \big\{ \bar{s}^a(x) \gamma^\mu Q^a(x) \bar{Q}(0) i \gamma^5 s^b(0) \big\} |0 \rangle_F 
\end{equation}
Moreover, the expressions of the heavy and light quark propagators in the presence of the background (gluonic and electromagnetic) fields are needed to calculate the perturbative part of the correlation function. While the heavy quark propagator is given by Eq.~\eqref{eq:7} the light quark propagator is
\begin{equation}
  \label{eq:25}
    S_q(x) = \frac{i \slashed{x}}{2 \pi^2 x^4} - \frac{i g_s}{16 \pi^2 x^2} \int_0^1 du \bigg\{\bar{u} \slashed{x} \sigma_{\alpha \beta} + u \sigma_{\alpha \beta} \slashed{x} \bigg\} G^{\alpha \beta}(ux) + ... ~.
\end{equation}
There are perturbative and non-perturbative contributions in this calculation. While the perturbative one is obtained when photon is radiated from light or heavy quark propagators, the non-perturbative contribution is due to the photon radiated from long distance. These contributions are obtained in the following way.

After applying the Fierz identities given in Eq.~\eqref{eq:8}, the following matrix elements describing the non-perturbative interaction of photon with quarks appear.
\begin{equation}
  \label{eq:26}
  \begin{split}
    \langle \gamma(q,\eta) | \bar{q} \Gamma_i q |0 \rangle \\
    \langle \gamma(q,\eta) | \bar{q} \Gamma_i G_{\lambda \tau}^{(l)} q | 0 \rangle
  \end{split}
\end{equation}
Here $\Gamma_i = \{ I, \gamma_5, \gamma_\alpha, i \gamma_\alpha \gamma_5, \sigma_{\alpha \beta}/\sqrt{2} \}$. These matrix elements are parameterized in terms of the photon distribution amplitudes (DA's) with definite twists and are obtained in~\cite{Ball:2002ps}. The theoretical part of the correlation function within the light cone sum rules is also calculated in~\cite{Rohrwild:2007yt,Li:2002kr,Gelhausen:2014jea,1996PhRvD..54..857A} and for completeness presented in Appendix C.

Similar to the sum rules for the strong coupling constant $g_2$, the sum rules for the coupling constants $f_2$ are obtained by matching the two representations of the correlation function and performing double Borel transformation over the variables $-p^2$ and $-(p+q)^2$ which suppress the excited states and continuum subtraction. This is achieved by using the quark-hadron duality.

Finally, we get the following sum rule for the coupling constant $f_2$ \\
\begin{equation}
  \begin{split}
    \Pi & =    - \frac{3}{4 \pi^2} \int_{(m_c+m_s)^2}^{s_0} ds~e^{-s/M^2} \times \\&  \bigg[
    -(e_c - e_s) (m_c - m_s) \lambda(1,a,b) + e_c m_c \ln \bigg( \frac{1+a-b+\lambda}{1+a-b-\lambda} \bigg) 
    + e_s m_s \ln \bigg( \frac{1-a+b+\lambda}{1-a+b-\lambda} \bigg) \bigg]  \\
    &+ \big( e^{ -m_c^2/M^2} - e^{-s_0/M^2} \big) \bigg[ e_s \frac{f_{3 \gamma} m_c \psi^a(u)}{2} + e_s M^2 \chi \langle \bar{s} s \rangle \varphi_\gamma(u)  - e_s \frac{\mathbb{A}(u)}{4} \langle \bar{s} s \rangle \big( 1 + \frac{m_c^2}{M^2} \big) \bigg] \\
    &+ I_1 + I_2 ~.
  \end{split}
\end{equation}
To derive this equation, we used $M_1^2 = M_2^2 = 2 M^2$. Here, $\chi$ is the magnetic susceptibility, $s_0$ is the continuum threshold, $a = m_c^2/s$, $b = m_s^2/s$, $\lambda(1,a,b) = (1 + a^2 + b^2 - 2 a b - 2a -2b)^{1/2}$ and the explicit form of the functions $I_1$ and $I_2$ can be found in~\cite{Rohrwild:2007yt}. Note that the results for $\BsprimetoBsstargamma$ transition is obtained with the help of the replacements $D_{s}(nS) \rightarrow B_{s}(nS)$, $D_s^* \rightarrow B_s^*$, $m_c \rightarrow m_b$, and $e_c \rightarrow e_b$.

We also would like to state the difference between our work with the ones in the literature~\cite{1996PhRvD..54..857A,Rohrwild:2007yt,Li:2002kr}.
\begin{itemize}
\item In \cite{Li:2002kr} and \cite{1996PhRvD..54..857A} the three particle distribution amplitudes for a photon which are presented in~\cite{Rohrwild:2007yt} are not taken into account.
  \item In~\cite{Rohrwild:2007yt,1996PhRvD..54..857A} the strange quark is neglected. 
  \end{itemize}

\section{Numerical Analysis}
\label{sec:3}
Here, we present the numerical analysis to determine the strong coupling $g_2$ and electromagnetic coupling $f_2$ obtained in the previous section. LCSR method contains numerous input parameters such as the light and heavy quark masses, the masses  of the corresponding mesons, decay constants of $D_{s}(nS)$ ($B_{s}(nS)$)  and $D^*(B^*)$ , $D_s^*(B_s^*)$ mesons, and the value of quark condensates.  These values are collected in Table~\ref{tab:1}. We used the $\overline{MS}$ values of the heavy quarks in our numerical analysis. Another sets of the input parameters are the DA's of K-meson and photon. These DA's are presented in~\cite{Ball:2006wn,Ball:1998sk}.

\begin{figure}[hbt!]
  \centering
  \includegraphics[scale=0.65]{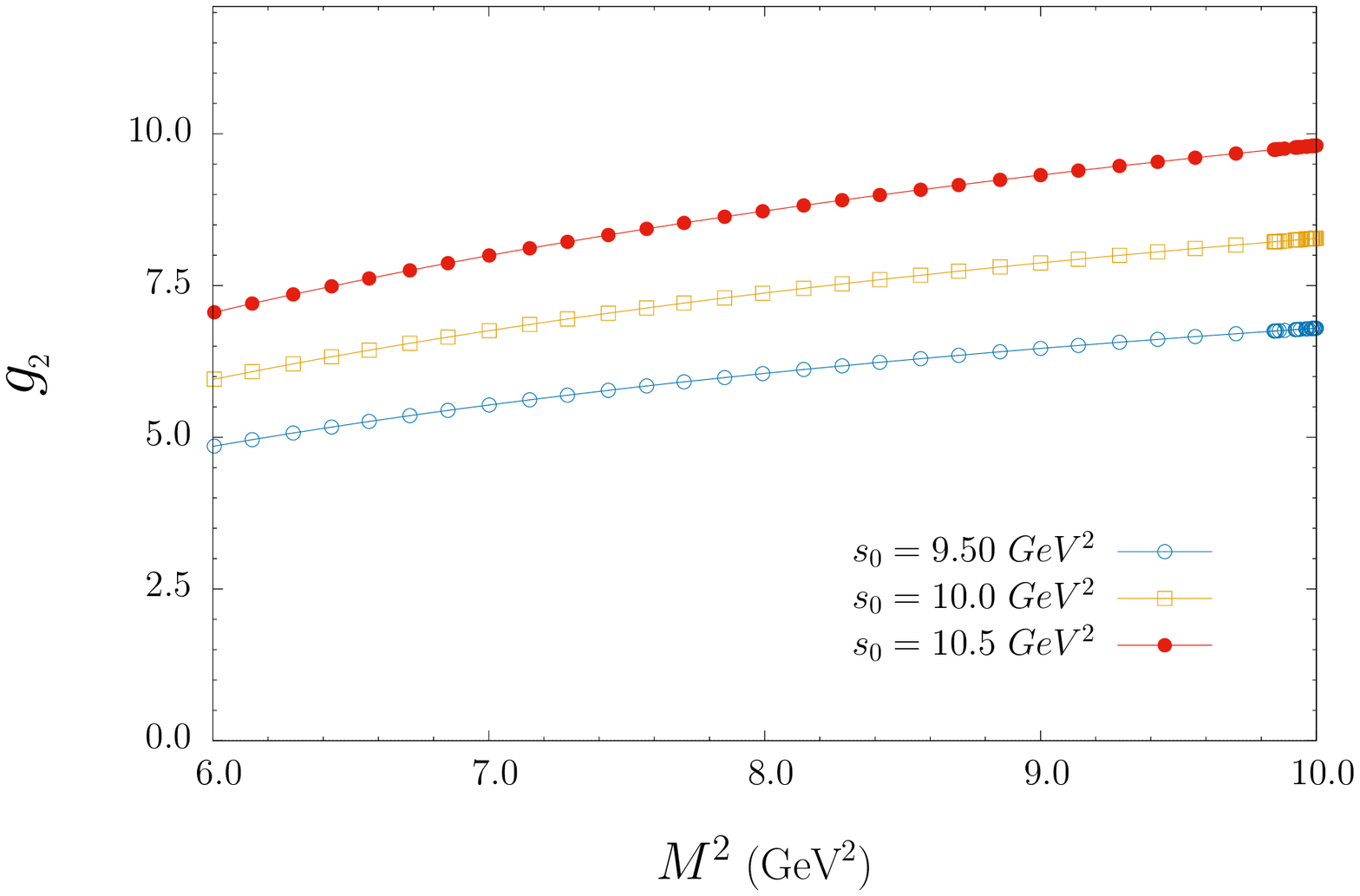}
  \caption{The dependency of $g$ on $M^2$ at the fixed values of $s_0$ for $\DsprimetoDstarK$ decay.}
  \label{fig:1}
\end{figure}

\begin{figure}[hbt!]
  \centering
  \includegraphics[scale=0.65]{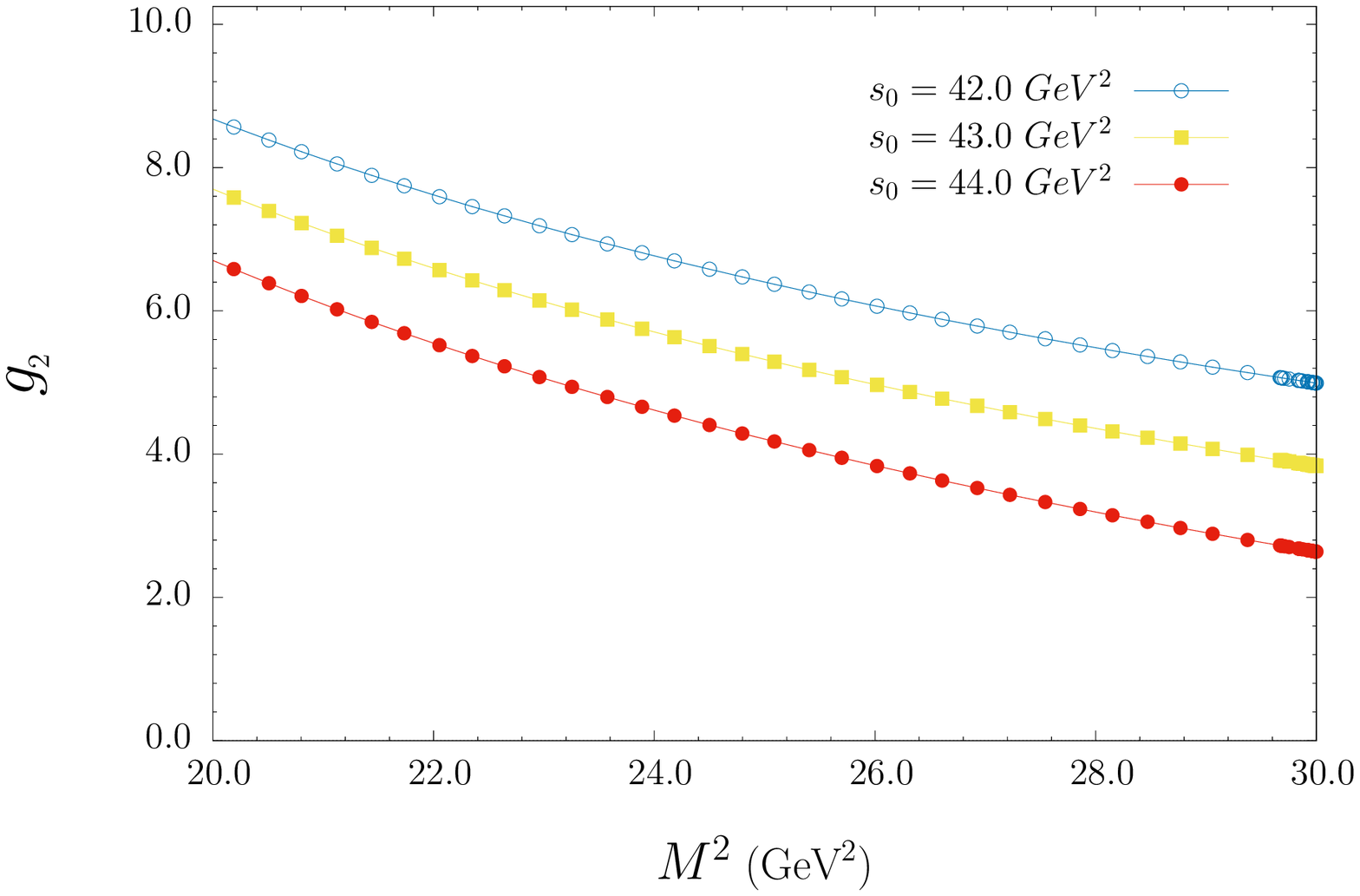}
  \caption{Same as in Fig.~\ref{fig:1} but for $\BsprimetoBstarK$ transition.}
  \label{fig:2}
\end{figure}

\begin{figure}[hbt!]
  \centering
  \includegraphics[scale=0.65]{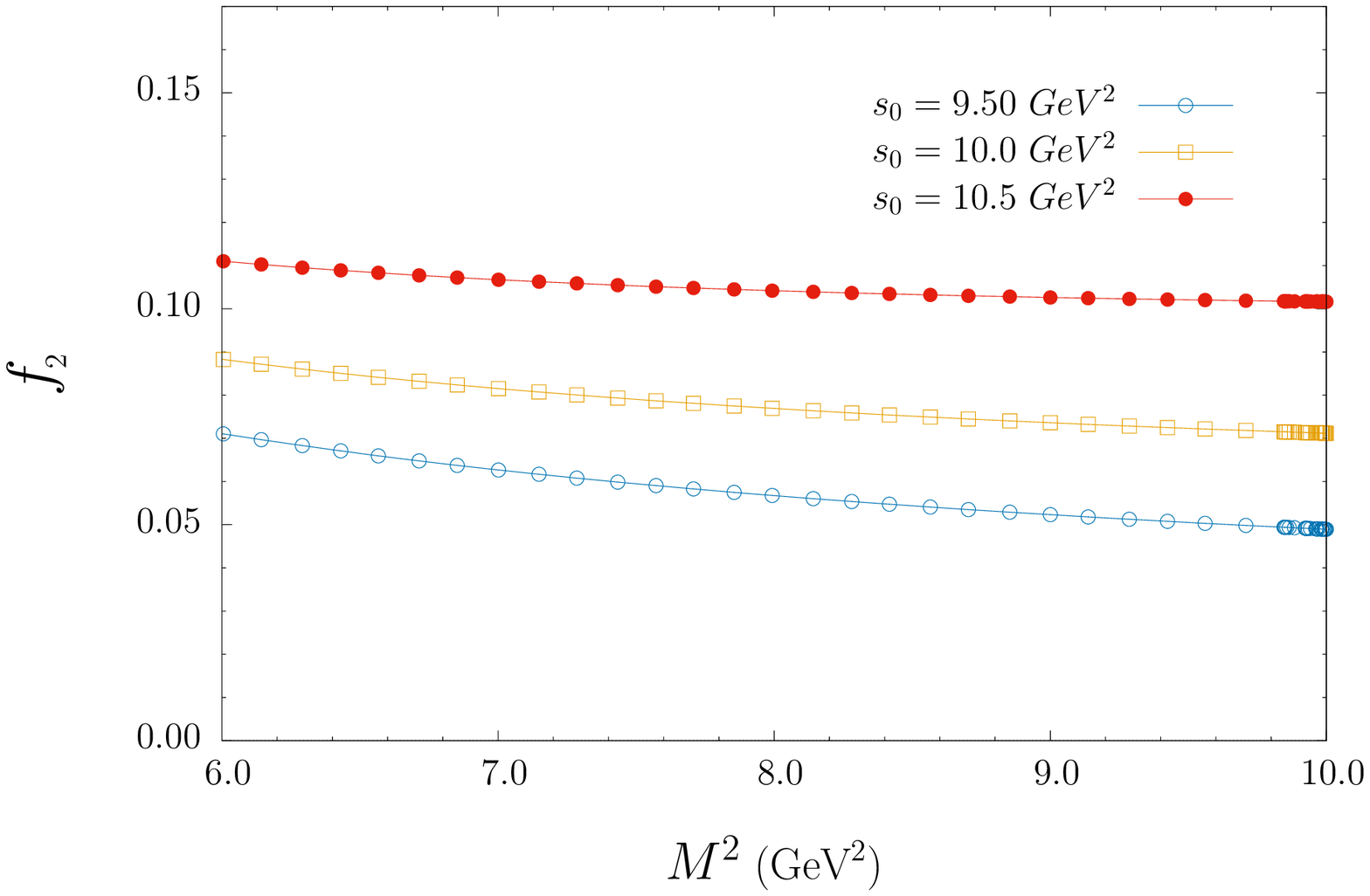}
  \caption{The dependency of $f_2 (GeV^{-1})$  on $M^2$ at fixed values of $s_0$ for $\DsprimetoDsstargamma$ }
  \label{fig:3}
\end{figure}

\begin{figure}[hbt!]
  \centering
  \includegraphics[scale=0.65]{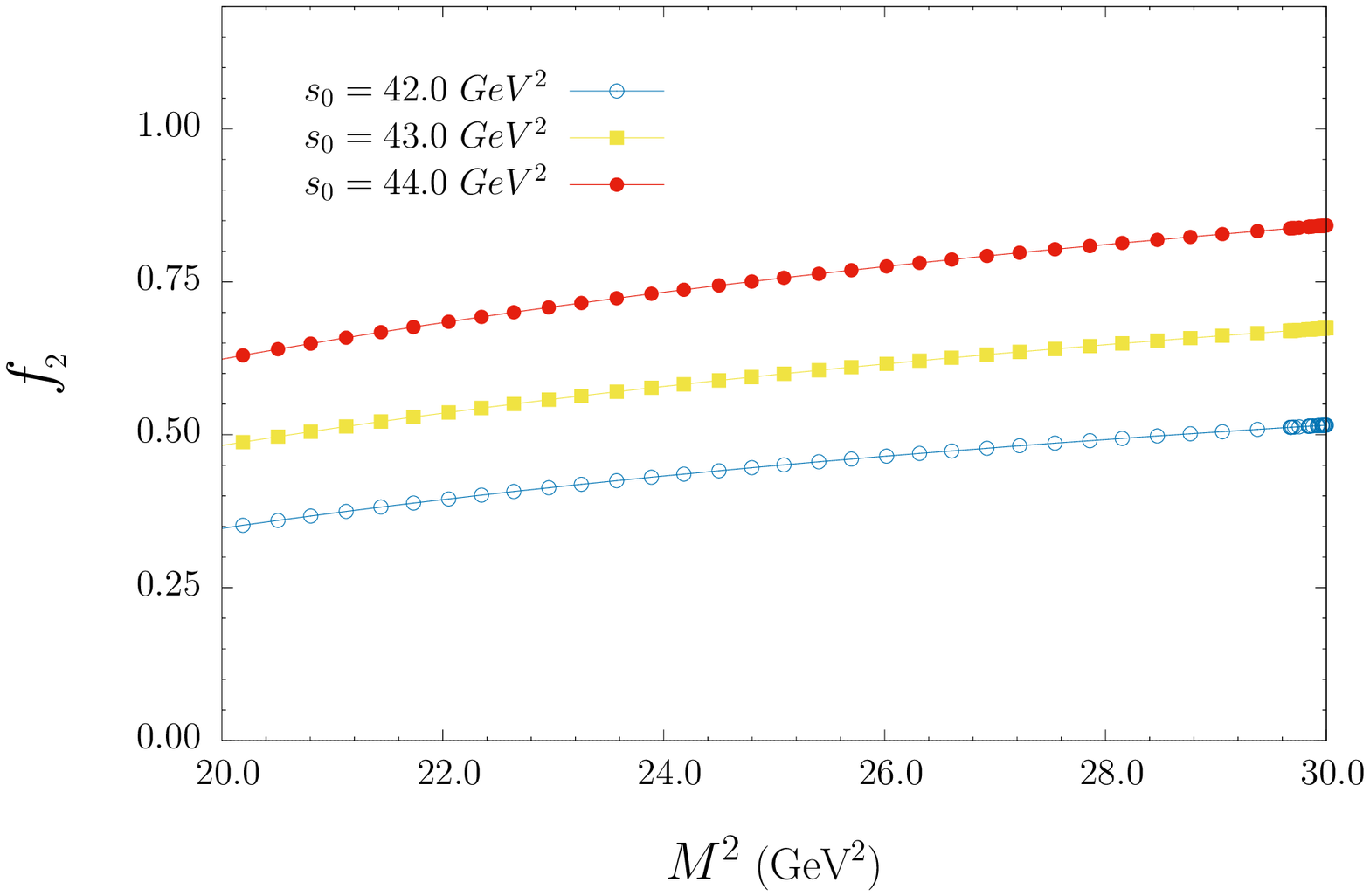}
  \caption{Same as in Fig.\ref{fig:3} but for $\BsprimetoBsstargamma$ transition.}
  \label{fig:4}
\end{figure}

In addition to these input parameters, the sum rules contain two auxiliary parameters, namely the Borel mass parameter, $M^2$, and continuum threshold $s_0$. Hence, we need to find the regions, so-called working regions, of these parameters where the couplings demonstrate weak dependency on the variation of these parameters.
\begin{table*}[hbt]
  \centering
  \renewcommand{\arraystretch}{1.4}
  \setlength{\tabcolsep}{7pt}
  \begin{tabular}{ccccc}
    \toprule
     Parameters             & Value         \\
    \midrule
    $\overline{m}_s(2 GeV)$ & $0.095 \pm 0.010~GeV$~\cite{PhysRevD.98.030001}\\ 
    $\overline{m}_c(\overline{m}_c)$ & $1.275 \pm 0.025~GeV$~\cite{PhysRevD.98.030001} \\ 
    $\overline{m}_b(\overline{m}_b)$ & $4.18 \pm 0.03$~\cite{PhysRevD.98.030001}  \\ 
    $m_K$    & $0.497~GeV$~\cite{PhysRevD.98.030001}  \\ 
    $\langle \bar{q} q \rangle$    & $-(0.245)^3~GeV^3$~\cite{PhysRevD.98.030001}  \\
    $\langle \bar{s} s \rangle$    & $(0.8 \pm 0.2)~\langle \bar{q} q \rangle ~GeV^3$ ~\cite{Gelhausen:2014jea}\\
    $f_{D_{s_2}}$ &  $0.143^{+0.019}_{-0.031}~GeV$~\cite{Gelhausen:2014jea}\\
    $f_{D_{s_1}}$ &  $0.279^{+0.021}_{-0.012}~GeV$~\cite{Gelhausen:2014jea}\\
    $f_{D_s^*}$ & $0.293^{+0.019}_{-0.014}~GeV$~\cite{Gelhausen:2014jea} \\
    $f_{D^*}$ & $0.235^{+0.025}_{-0.012}~GeV$~\cite{Gelhausen:2014jea} \\
    $f_{B_{s_2}}$ & $0.174^{+0.019}_{-0.019}~GeV$~\cite{Gelhausen:2014jea}\\
    $f_{B_{s_1}}$ & $0.244^{+0.013}_{-0.026}~GeV$~\cite{Gelhausen:2014jea}\\
    $f_{B_s^*}$ & $0.251^{+0.014}_{-0.016}~GeV$ ~\cite{Gelhausen:2014jea}\\
    $f_{B^*}$ & $0.208^{+0.012}_{-0.021}~GeV$ ~\cite{Gelhausen:2014jea}\\
    $f_{3 \gamma}$ & $ - (4 \pm 2) \times 10^{-3} ~GeV^2$ ~\cite{Rohrwild:2007yt}\\
    $ \chi( \mu = 1~GeV)$ & $3.15 \pm 0.3 ~GeV^{-2}$ ~\cite{Rohrwild:2007yt} \\
    \bottomrule
  \end{tabular}
  \caption{The values of the input parameters used in our calculations.}
  \label{tab:1}
\end{table*}
The lowest value of $M^2$ is obtained by requiring the condition that the higher twists contributions should considerably be smaller than the lowest twists terms and constitute maximum $15 \%$ of the contribution. The upper bound on $M^2$ is determined by demanding that the continuum contribution be less than $30 \%$ of the pole contribution. Considering these requirements, we find  the following working regions of $M^2$
\begin{equation}
  \label{eq:10}
  \begin{split}
    & 6~GeV^2 \leq M^2 \leq 10 ~GeV^2   \hspace{0.8cm} \text{for } \DsprimetoDstarK \text{ and } \DsprimetoDsstargamma, \\
    & 20~GeV^2 \leq M^2 \leq 30 ~GeV^2 \hspace{0.6cm} \text{for } \BsprimetoBstarK \text{ and } \BsprimetoBsstargamma~. 
  \end{split}
\end{equation}
The working region of the continuum threshold $s_0$ is determined by requiring that the two-point sum rules reproduce a $10\%$ accuracy of the mass of the radially excited states. With this restriction, we obtain the following optimum values:
\begin{equation}
  \label{eq:11}
  \begin{split}
    & 9.5~GeV^2 \leq s_0 \leq 10.5 ~GeV^2   \hspace{0.6cm} \text{for } \DsprimetoDstarK \text{ and } \DsprimetoDsstargamma, \\
    & 42~GeV^2 \leq s_0 \leq 44 ~GeV^2      \hspace{1.0cm} \text{for } \BsprimetoBstarK \text{ and } \BsprimetoBsstargamma~. 
  \end{split}
\end{equation}
%
In Figs.~\ref{fig:1} and \ref{fig:2}, we present the dependencies of $g_2$ on $M^2$ at the fixed values of $s_0$ for $\DsprimetoDstarK$ and $\BsprimetoBstarK$, respectively. From these figures it follows that the coupling constant shows good stability for the regions $6~GeV^2 \leq M^2 \leq 10~GeV^2$ and $22~GeV^2 \leq M^2 \leq 30~GeV^2$, respectively. Taking into account all the uncertainties in the values of the input parameters we get:
\begin{equation}
  \label{eq:13}
  \begin{split}
    g_2 & = (7.5 \pm 1.6) \hspace{1.5cm} \text{for } D_{s}(2S) \rightarrow D^* K,  \\
    g_2 & = (4.6 \pm 1.3) \hspace{1.5cm} \text{for } B_{s}(2S) \rightarrow B^* K~.
  \end{split}
\end{equation}
In Figs.~\ref{fig:3} and ~\ref{fig:4} we depict the dependency of the coupling constant $f_2$ on $M^2$ at fixed values of $s_0$ for $\DsprimetoDsstargamma$ and $\BsprimetoBsstargamma$, respectively. Once we take into account the uncertainties in the input parameters, we obtain the following values for $f_2$:
\begin{equation}
  \label{eq:14}
  \begin{split}
    f_2 & = (0.08 \pm 0.02) ~GeV^{-1} \hspace{1.25cm} \text{for } D_{s}(2S) \rightarrow D_s^* \gamma  \\
    f_2 & = (0.58 \pm 0.22) ~GeV^{-1} \hspace{1.25cm} \text{for } B_{s}(2S) \rightarrow B_s^* \gamma
  \end{split}
\end{equation}

The obtained values of $g_2$ and $f_2$ allowed us to predict the corresponding decay widths via following formulas
\begin{equation}
  \label{eq:14}
  \begin{split}
    \Gamma({D_{s}(2S)} \rightarrow {D_{s}^*} K) &= \frac{g_2^2}{64 \pi }
    \frac{ \big((m_{D_{s}(2S)} - m_{D^*})^2 - m_{K}^2 \big)^{3/2} \big((m_{D_{s}(2S)} + m_{D^*})^2 - m_{K}^2 \big)^{3/2}}{m_{D_{s}(2S)}^5}, \\
    \Gamma({D_{s}(2S)} \rightarrow {D_{s}^*} \gamma) &= \frac{f_2^2}{32 \pi} \frac{ (m_{D_{s}(2S)}^2 - m_{D_{s}^*}^2)^3}{ m_{D_{s}(2S)}^3} ,
  \end{split}
\end{equation}
%
%
Using the values of $g_2$ and $f_2$, we obtain the following values for the corresponding decay widths
\begin{equation}
  \label{eq:15}
  \begin{split}
    \Gamma (\DsprimetoDstarK) &= (6.14 \pm 2.62)~MeV \\
    \Gamma (\BsprimetoBstarK) &= (3.11 \pm 1.77)~MeV \\
    \Gamma (\DsprimetoDsstargamma) &= (42 \pm 21)~keV \\
    \Gamma (\BsprimetoBsstargamma) &= (6.23 \pm 4.62)~MeV 
  \end{split}
\end{equation}
All the errors from different sources are taken into account quadratically in this study.

We see from these results that the strong decay widths of the radially excited $D_s$ and $B_s$ mesons are quite large and can be potentially observed at LHCb. In addition, the radiative decay widths are also larger than $D_s^*(1S) \rightarrow D_s(1S) \gamma$ and $B_s^*(1S) \rightarrow B_s(1S) \gamma$ can be observed in future experiments.

At the end of this section, we would like to make the following remarks. The $g_{D^* D \pi}$ and $g_{B^* B \pi}$ couplings without and with NLO corrections are calculated in~\cite{Khodjamirian:2020mlb,PhysRevD.51.6177}, respectively. It is obtained that the NLO corrections increase the results by around $12 \%$ and $4 \%$ in D and B mesons sectors, respectively. We expect that the NLO corrections would alter the results in the same order. 
%
%
%
\section{Conclusion}
The discovery of new radially excited $D_{s}(2S)$ and $B_{s}(2S)$ mesons at LHCb stimulated theoretical and experimental studies for a deeper understanding of the properties of these mesons and the radial excitation mesons in general. We estimated these mesons' most promising decay channels within the light cone QCD sum rules method in the present work. From the obtained results, one can conclude that these decays would have a chance to be observed in future experiments.
\section{Acknowledgment}
The author thanks T.M.Aliev and M.Savci for valuable discussions. 

\clearpage


\bibliographystyle{utcaps_mod}
\bibliography{all.bib}



\section*{Appendix A}
\setcounter{equation}{0}
\setcounter{section}{0}

The matrix elements of the nonlocal
operators between vacuum and one--particle light pseudoscalar meson states
are needed.
The matrix elements $\lla {\cal P}(q) \vel \bar{q} (x) \Gamma_i q^\prime (0) \ver 0
\rra$ are parametrized in terms of the DA's
\cite{Ball:2006wn,Ball:2004ye,Ball:1998sk} and they are determined as,
\bea
\label{eozd15} 
\lla {\cal P}(p)\vel \bar q_1(x) \gamma_\mu \gamma_5 q_1(0)\ver 0 \rra \es
-i f_{\cal P} q_\mu  \int_0^1 du  e^{i \bar u q x}
    \left( \varphi_{\cal P}(u) + {1\over 16} m_{\cal P}^2
x^2 {\Bbb{A}}(u) \right) \nnb \\
\ek {i\over 2} f_{\cal P} m_{\cal P}^2 {x_\mu\over qx}
\int_0^1 du e^{i \bar u qx} {\Bbb{B}}(u)~,\nnb \\
\lla {\cal P}(p)\vel \bar q_1(x) i \gamma_5 q_2(0)\ver 0 \rra \es
\mu_{\cal P} \int_0^1 du e^{i \bar u qx} \phi_P(u)~,\nnb \\
\lla {\cal P}(p)\vel \bar q_1(x) \sigma_{\alpha \beta} \gamma_5 q_2(0)\ver 0 \rra \es
{i\over 6} \mu_{\cal P} \left( 1 - \widetilde{\mu}_{\cal P}^2 \right)
\left( q_\alpha x_\beta - q_\beta x_\alpha\right)
\int_0^1 du e^{i \bar u qx} \phi_\sigma(u)~,\nnb \\
\lla {\cal P}(p)\vel \bar q_1(x) \sigma_{\mu \nu} \gamma_5 g_s
G_{\alpha \beta}(v x) q_2(0)\ver 0 \rra \es i \mu_{\cal P} \left[
q_\alpha q_\mu \left( g_{\nu \beta} - {1\over qx}(q_\nu x_\beta +
q_\beta x_\nu) \right) \right. \nnb \\
\ek q_\alpha q_\nu \left( g_{\mu \beta} -
{1\over qx}(q_\mu x_\beta + q_\beta x_\mu) \right) \nnb \\
\ek q_\beta q_\mu \left( g_{\nu \alpha} - {1\over qx}
(q_\nu x_\alpha + q_\alpha x_\nu) \right) \nnb \\
\ar q_\beta q_\nu \left. \left( g_{\mu \alpha} -
{1\over qx}(q_\mu x_\alpha + q_\alpha x_\mu) \right) \right] \nnb \\
\cp \int {\cal D} \alpha e^{i (\alpha_{\bar q} +
v \alpha_g) qx} {\cal T}(\alpha_i)~,\nnb \\
\lla {\cal P}(p)\vel \bar q_1(x) \gamma_\mu \gamma_5 g_s
G_{\alpha \beta} (v x) q_2(0)\ver 0 \rra \es q_\mu (q_\alpha x_\beta -
q_\beta x_\alpha) {1\over qx} f_{\cal P} m_{\cal P}^2
\int {\cal D}\alpha e^{i (\alpha_{\bar q} + v \alpha_g) qx}
{\cal A}_\parallel (\alpha_i) \nnb \\
\ar \left[q_\beta \left( g_{\mu \alpha} - {1\over qx}
(q_\mu x_\alpha + q_\alpha x_\mu) \right) \right. \nnb \\
\ek q_\alpha \left. \left(g_{\mu \beta}  - {1\over qx}
(q_\mu x_\beta + q_\beta x_\mu) \right) \right]
f_{\cal P} m_{\cal P}^2 \nnb \\
\cp \int {\cal D}\alpha e^{i (\alpha_{\bar q} + v \alpha _g)
q x} {\cal A}_\perp(\alpha_i)~,\nnb \\
\lla {\cal P}(p)\vel \bar q_1(x) \gamma_\mu i g_s G_{\alpha \beta}
(v x) q_2(0)\ver 0 \rra \es q_\mu (q_\alpha x_\beta - q_\beta x_\alpha)
{1\over qx} f_{\cal P} m_{\cal P}^2 \int {\cal D}\alpha e^{i (\alpha_{\bar q} +
v \alpha_g) qx} {\cal V}_\parallel (\alpha_i) \nnb \\
\ar \left[q_\beta \left( g_{\mu \alpha} - {1\over qx}
(q_\mu x_\alpha + q_\alpha x_\mu) \right) \right. \nnb \\
\ek q_\alpha \left. \left(g_{\mu \beta}  - {1\over qx}
(q_\mu x_\beta + q_\beta x_\mu) \right) \right] f_{\cal P} m_{\cal P}^2 \nnb \\
\cp \int {\cal D}\alpha e^{i (\alpha_{\bar q} +
v \alpha _g) q x} {\cal V}_\perp(\alpha_i)~,
\eea
where
\bea
\label{nolabel07}
\mu_{\cal P} = f_{\cal P} {m_{\cal P}^2\over m_{q_1} + m_{q_2}}~,~~~~~
\widetilde{\mu}_{\cal P} = {m_{q_1} + m_{q_2} \over m_{\cal P}}~, \nnb
\eea
and $q_1$ and $q_2$ are the quarks in the meson ${\cal P}$,
${\cal D}\alpha = d\alpha_{\bar q} d\alpha_q d\alpha_g
\delta(1-\alpha_{\bar q} - \alpha_q - \alpha_g)$.
Here 
$\varphi_{\cal P}(u)$ is the leading twist--two, $\phi_P(u)$, $\phi_\sigma(u)$,
${\cal T}(\alpha_i)$ are the twist--three, and
$\Bbb{A}(u)$, $\Bbb{B}(u)$, ${\cal A}_\perp(\alpha_i),$ ${\cal A}_\parallel(\alpha_i),$
${\cal V}_\perp(\alpha_i)$ and ${\cal V}_\parallel(\alpha_i)$
are the twist--four DAs, respectively. 

The main input parameters of the light cone QCD sum rules are the
distribution amplitudes, whose expressions are given below
\cite{Ball:2006wn,Ball:2004ye,Ball:1998sk},

\bea
\label{eozd17}
\varphi_{\cal P}(u) \es 6 u \bar u \left[ 1 + a_1^{\cal P} C_1(2 u -1) +
a_2^{\cal P} C_2^{3/2}(2 u - 1) \right]~,  \nnb \\
{\cal T}(\alpha_i) \es 360 \eta_3 \alpha_{\bar q} \alpha_q
\alpha_g^2 \left[ 1 + w_3 {1\over 2} (7 \alpha_g-3) \right]~, \nnb \\
\phi_P(u) \es 1 + \left[ 30 \eta_3 - {5\over 2}
{1\over \mu_{\cal P}^2}\right] C_2^{1/2}(2 u - 1)~,  \nnb \\
\ar \left( -3 \eta_3 w_3  - {27\over 20} {1\over \mu_{\cal P}^2} -
{81\over 10} {1\over \mu_{\cal P}^2} a_2^{\cal P} \right)
C_4^{1/2}(2u-1)~, \nnb \\
\phi_\sigma(u) \es 6 u \bar u \left[ 1 + \left(5 \eta_3 - {1\over 2} \eta_3 w_3 -
{7\over 20}  \mu_{\cal P}^2 - {3\over 5} \mu_{\cal P}^2 a_2^{\cal P} \right)
C_2^{3/2}(2u-1) \right] ~, \nnb \\
{\cal V}_\parallel(\alpha_i) \es 120 \alpha_q \alpha_{\bar q} \alpha_g
\left( v_{00} + v_{10} (3 \alpha_g -1) \right) ~, \nnb \\
{\cal A}_\parallel(\alpha_i) \es 120 \alpha_q \alpha_{\bar q} \alpha_g
\left( 0 + a_{10} (\alpha_q - \alpha_{\bar q}) \right) ~, \nnb \\
{\cal V}_\perp (\alpha_i) \es - 30 \alpha_g^2\left[ h_{00}(1-\alpha_g) +
h_{01} (\alpha_g(1-\alpha_g)- 6 \alpha_q \alpha_{\bar q}) +
h_{10}(\alpha_g(1-\alpha_g) - {3\over 2} (\alpha_{\bar q}^2+
\alpha_q^2)) \right] ~, \nnb \\
{\cal A}_\perp (\alpha_i) \es 30 \alpha_g^2(\alpha_{\bar q} - \alpha_q)
\left[ h_{00} + h_{01} \alpha_g + {1\over 2} h_{10}(5 \alpha_g-3) \right] ~, \nnb \\
B(u)\es g_{\cal P}(u) - \varphi_{\cal P}(u) ~, \nnb \\
g_{\cal P}(u) \es g_0 C_0^{1/2}(2 u - 1) + g_2 C_2^{1/2}(2 u - 1) +
g_4 C_4^{1/2}(2 u - 1) ~, \nnb \\
\Bbb{A}(u) \es 6 u \bar u \left[{16\over 15} + {24\over 35} a_2^{\cal P}+
20 \eta_3 + {20\over 9} \eta_4 +
\left( - {1\over 15}+ {1\over 16}- {7\over 27}\eta_3 w_3 -
{10\over 27} \eta_4 \right) C_2^{3/2}(2 u - 1)  \right. \nnb \\
    \ar \left. \left( - {11\over 210}a_2^{\cal P} - {4\over 135}
\eta_3w_3 \right)C_4^{3/2}(2 u - 1)\right] ~, \nnb \\
\ar \left( -{18\over 5} a_2^{\cal P} + 21 \eta_4 w_4 \right)
\left[ 2 u^3 (10 - 15 u + 6 u^2) \ln u  \right. \nnb \\
\ar \left. 2 \bar u^3 (10 - 15 \bar u + 6 \bar u ^2) \ln\bar u +
u \bar u (2 + 13 u \bar u) \right]~,
\eea
where $C_n^k(x)$ are the Gegenbauer polynomials, and
\bea
\label{eozd18}
h_{00}\es v_{00} = - {1\over 3}\eta_4 ~, \nnb \\
a_{10} \es {21\over 8} \eta_4 w_4 - {9\over 20} a_2^{\cal P} ~, \nnb \\
v_{10} \es {21\over 8} \eta_4 w_4 ~, \nnb \\
h_{01} \es {7\over 4}  \eta_4 w_4  - {3\over 20} a_2^{\cal P} ~, \nnb \\
h_{10} \es {7\over 4} \eta_4 w_4 + {3\over 20} a_2^{\cal P} ~, \nnb \\
g_0 \es 1 ~, \nnb \\
g_2 \es 1 + {18\over 7} a_2^{\cal P} + 60 \eta_3  + {20\over 3} \eta_4 ~, \nnb \\
g_4 \es  - {9\over 28} a_2^{\cal P} - 6 \eta_3 w_3~.
\eea
The values of the parameters $a_1^{\cal P}$, $a_2^{\cal P}$,
$\eta_3$, $\eta_4$, $w_3$, and $w_4$ entering Eq. (\ref{eozd18}) are listed in
Table (\ref{param}) for the pseudoscalar $\pi$, $K$ and $\eta$ mesons.

\begin{table}[h]
\def\bos{\lower 0.25cm\hbox{{\vrule width 0pt height 0.7cm}}}
\begin{center}
\begin{tabular}{|c|c|c|}
\hline\hline
        & \bos  $\pi$   &   $K$ \\
\hline
$a_1^{\cal P}$  & \bos   0 &   0.050 \\
\hline
$a_2^{\cal P}~\mbox{(set-1)}$  & \bos   0.11  &   0.15 \\
\hline
$a_2^{\cal P}~\mbox{(set-2)}$  &  \bos  0.25  &   0.27 \\
\hline
$\eta_3$    & \bos  0.015 &   0.015 \\
\hline
$\eta_4$    & \bos  10    &   0.6 \\
\hline
$w_3$       & \bos  $-3$    &   $-3$ \\
\hline
$w_4$       & \bos  0.2   &   0.2 \\
\hline \hline
\end{tabular}
\end{center}
\caption{Parameters of the wave function calculated at the renormalization scale $\mu = 1 ~GeV$}
\label{param}
\end{table}

\section*{Appendix B}
\label{sec:appendix}
\setcounter{equation}{0}
\subsection*{ Coefficient of the
$q_{\mu}$ structure for  $\DsprimetoDstarK$  decay}
\bea
\Pi \es
%
%
{1\over 12} M^2 \Big\{24 \mu_{\cal P} i_3({\cal T},v) + 12 f_{\cal P} m_Q \varphi_{\cal P}(\bar{u}_0) +
6 \mu_{\cal P} \phi_P(\bar{u}_0) \nnb \\
\ar \mu_{\cal P} (1 - \widetilde{\mu}_{\cal P}^2) \Big[4 \phi_{\sigma}(\bar{u}_0) -
\phi_{\sigma}^{\prime}(\bar{u}_0) \Big] \Big\} \nnb \\
%
%
\ar {1\over 24 M^2}
\Big\{f_{\cal P} m_Q (-6 m_{\cal P}^2 m_Q^2  \mathbb{A}(\bar{u}_0) -
24 m_{\cal P}^4 \Big[i_1({\cal A}_{\parallel},1) + i_1({\cal A}_{\perp},1) -
i_1({\cal V}_{\parallel},1) - i_1({\cal V}_{\perp},1) \Big] \nnb \\
\ar \GG \varphi_{\cal P}(\bar{u}_0))\Big\} \nnb \\
%
%
\ek  {1 \over 864 M^4}
\Big\{m_Q \GG \Big[-18 f_{\cal P} m_{\cal P}^2  \mathbb{A}(\bar{u}_0) +
18 f_{\cal P} m_{\cal P}^2 \widetilde{j}_1(B) \nnb \\
\ar m_Q \Big(12 f_{\cal P} m_Q \varphi_{\cal P}(\bar{u}_0) +
6 \mu_{\cal P} \phi_P(\bar{u}_0) -
\mu_{\cal P} (1 - \widetilde{\mu}_{\cal P}^2) \{8 \phi_{\sigma}(\bar{u}_0) +
\phi_{\sigma}^\prime(\bar{u}_0)\}]\Big) \Big]\Big\} \nnb \\
%
%
\ek {1\over 432 M^6}
\Big\{m_Q^2 \GG \Big[3 f_{\cal P} m_{\cal P}^2 m_Q \Big(3  \mathbb{A}(\bar{u}_0) -
\widetilde{j}_1(B) \Big) - (m_{\cal P}^2 - 2 m_Q^2) \mu_{\cal P}
(1 - \widetilde{\mu}_{\cal P}^2) \phi_{\sigma}(\bar{u}_0) \Big] \Big\} \nnb \\
%
%
\ar {1 \over 288 M^8}
\Big[f_{\cal P} m_{\cal P}^2 m_Q^5 \GG  \mathbb{A}(\bar{u}_0) \Big] \nnb \\
%
%
\ek {1\over 2}\Big[f_{\cal P} m_{\cal P}^2 m_Q \widetilde{j}_1(B)\Big] \nnb \\
\ek {1\over 6} \Big\{3 m_{\cal P}^2 \Big[3 \mu_{\cal P} i_2({\cal T},1)
+ 2 f_{\cal P} m_Q \Big(i_2({\cal A}_{\parallel},1) -
2 i_2({\cal A}_{\perp},1) - i_2({\cal V}_{\parallel},1) + 
2 i_2({\cal V}_{\perp},1) \Big)\nnb \\
\ek 10 \mu_{\cal P}  i_2({\cal T},v) \Big]
+(m_{\cal P}^2 - 2 m_Q^2) \mu_{\cal P} 
(1 - \widetilde{\mu}_{\cal P}^2) \phi_\sigma(\bar{u}_0)
\Big\}~.
\eea
where
\bea
\label{nolabel}
\mu_{\cal P} \es {f_{\cal P}
m_{\cal P}^2 \over m_{q_1}+m_{q_2}}~,~~~~~
\widetilde{\mu}_{\cal P} = {m_{q_1}+m_{q_2} \over m_{\cal P}}~,
\eea

The functions $i_n~(n=1,2)$, and $\widetilde{j}_1(f(u))$
are defined as:
\bea
\label{nolabel}
i_1(\phi,f(v)) \es \int {\cal D}\alpha_i \int_0^1 dv
\phi(\alpha_{\bar{q}},\alpha_q,\alpha_g) f(v) \delta(k-u_0)~, \nnb \\
i_2(\phi,f(v)) \es \int {\cal D}\alpha_i \int_0^1 dv
\phi(\alpha_{\bar{q}},\alpha_q,\alpha_g) f(v) \delta^\prime(k-u_0)~, \nnb \\
\widetilde{j}_1(f(u)) \es \int_{u_0}^1 du f(u)~, \nnb \\
{\cal I}_n \es \int_{m_Q^2}^{\infty} ds\, {e^{-s/M^2} \over s^n}~,\nnb \\
{\cal I}_n^{\ell n}  \es \int_{m_Q^2}^{\infty} ds\, {e^{-s/M^2} \over s^n} 
\ln{M^2 (s-m_Q^2) \over \Lambda^2 s}~,\nnb \\
\eea

where 
\bea
k = \alpha_q + \alpha_g (1-v)~,~~~
\widetilde{k} = \alpha_{\bar{q}} + \alpha_g v~,~~~
u_0={M_1^2 \over M_1^2
+M_2^2}~,~~~{\bar u}_0 = 1 - u_0~,~~~M^2={M_1^2 M_2^2 \over M_1^2 +M_2^2}~.\nnb
\eea


\section*{Appendix C}
\setcounter{equation}{0}
\setcounter{section}{0}

For completeness, in this Appendix we present the matrix elements  $\langle \gamma(q) | \bar{q} \Gamma_i q | 0 \rangle $ and $\langle \gamma(q) | \bar{q} \Gamma_i G_{\mu \nu} q | 0 \rangle $ which are calculated in terms of the photon DA's \cite{Ball:2002ps}.

 \bea
&&\langle \gamma(q) \vert  \bar q(x) \sigma_{\mu \nu} q(0) \vert  0
\rangle  = -i e_q \bar q q (\varepsilon_\mu q_\nu - \varepsilon_\nu
q_\mu) \int_0^1 du e^{i \bar u qx} \left(\chi \varphi_\gamma(u) +
\frac{x^2}{16} \mathbb{A}  (u) \right) \nnb \\ &&
-\frac{i}{2(qx)}  e_q \qq \left[x_\nu \left(\varepsilon_\mu - q_\mu
\frac{\varepsilon x}{qx}\right) - x_\mu \left(\varepsilon_\nu -
q_\nu \frac{\varepsilon x}{q x}\right) \right] \int_0^1 du e^{i \bar
u q x} h_\gamma(u)
\nnb \\
&&\langle \gamma(q) \vert  \bar q(x) \gamma_\mu q(0) \vert 0 \rangle
= e_q f_{3 \gamma} \left(\varepsilon_\mu - q_\mu \frac{\varepsilon
x}{q x} \right) \int_0^1 du e^{i \bar u q x} \psi^v(u)
\nnb \\
&&\langle \gamma(q) \vert \bar q(x) \gamma_\mu \gamma_5 q(0) \vert 0
\rangle  = - \frac{1}{4} e_q f_{3 \gamma} \epsilon_{\mu \nu \alpha
\beta } \varepsilon^\nu q^\alpha x^\beta \int_0^1 du e^{i \bar u q
x} \psi^a(u)
\nnb \\
&&\langle \gamma(q) | \bar q(x) g_s G_{\mu \nu} (v x) q(0) \vert 0
\rangle = -i e_q \qq \left(\varepsilon_\mu q_\nu - \varepsilon_\nu
q_\mu \right) \int {\cal D}\alpha_i e^{i (\alpha_{\bar q} + v
\alpha_g) q x} {\cal S}(\alpha_i)
\nnb \\
&&\langle \gamma(q) | \bar q(x) g_s \tilde G_{\mu \nu} i \gamma_5 (v
x) q(0) \vert 0 \rangle = -i e_q \qq \left(\varepsilon_\mu q_\nu -
\varepsilon_\nu q_\mu \right) \int {\cal D}\alpha_i e^{i
(\alpha_{\bar q} + v \alpha_g) q x} \tilde {\cal S}(\alpha_i)
\nnb \\
&&\langle \gamma(q) \vert \bar q(x) g_s \tilde G_{\mu \nu}(v x)
\gamma_\alpha \gamma_5 q(0) \vert 0 \rangle = e_q f_{3 \gamma}
q_\alpha (\varepsilon_\mu q_\nu - \varepsilon_\nu q_\mu) \int {\cal
D}\alpha_i e^{i (\alpha_{\bar q} + v \alpha_g) q x} {\cal
A}(\alpha_i)
\nnb \\
&&\langle \gamma(q) \vert \bar q(x) g_s G_{\mu \nu}(v x) i
\gamma_\alpha q(0) \vert 0 \rangle = e_q f_{3 \gamma} q_\alpha
(\varepsilon_\mu q_\nu - \varepsilon_\nu q_\mu) \int {\cal
D}\alpha_i e^{i (\alpha_{\bar q} + v \alpha_g) q x} {\cal
V}(\alpha_i) \nnb \\ && \langle \gamma(q) \vert \bar q(x)
\sigma_{\alpha \beta} g_s G_{\mu \nu}(v x) q(0) \vert 0 \rangle  =
e_q \qq \left\{
        \left[\left(\varepsilon_\mu - q_\mu \frac{\varepsilon x}{q x}\right)\left(g_{\alpha \nu} -
        \frac{1}{qx} (q_\alpha x_\nu + q_\nu x_\alpha)\right) \right. \right. q_\beta
\nnb \\ && -
         \left(\varepsilon_\mu - q_\mu \frac{\varepsilon x}{q x}\right)\left(g_{\beta \nu} -
        \frac{1}{qx} (q_\beta x_\nu + q_\nu x_\beta)\right) q_\alpha
\nnb \\ && -
         \left(\varepsilon_\nu - q_\nu \frac{\varepsilon x}{q x}\right)\left(g_{\alpha \mu} -
        \frac{1}{qx} (q_\alpha x_\mu + q_\mu x_\alpha)\right) q_\beta
\nnb \\ &&+
         \left. \left(\varepsilon_\nu - q_\nu \frac{\varepsilon x}{q.x}\right)\left( g_{\beta \mu} -
        \frac{1}{qx} (q_\beta x_\mu + q_\mu x_\beta)\right) q_\alpha \right]
   \int {\cal D}\alpha_i e^{i (\alpha_{\bar q} + v \alpha_g) qx} {\cal T}_1(\alpha_i)
\nnb \\ &&+
        \left[\left(\varepsilon_\alpha - q_\alpha \frac{\varepsilon x}{qx}\right)
        \left(g_{\mu \beta} - \frac{1}{qx}(q_\mu x_\beta + q_\beta x_\mu)\right) \right. q_\nu
\nnb \\ &&-
         \left(\varepsilon_\alpha - q_\alpha \frac{\varepsilon x}{qx}\right)
        \left(g_{\nu \beta} - \frac{1}{qx}(q_\nu x_\beta + q_\beta x_\nu)\right)  q_\mu
\nnb \\ && -
         \left(\varepsilon_\beta - q_\beta \frac{\varepsilon x}{qx}\right)
        \left(g_{\mu \alpha} - \frac{1}{qx}(q_\mu x_\alpha + q_\alpha x_\mu)\right) q_\nu
\nnb \\ &&+
         \left. \left(\varepsilon_\beta - q_\beta \frac{\varepsilon x}{qx}\right)
        \left(g_{\nu \alpha} - \frac{1}{qx}(q_\nu x_\alpha + q_\alpha x_\nu) \right) q_\mu
        \right]
    \int {\cal D} \alpha_i e^{i (\alpha_{\bar q} + v \alpha_g) qx} {\cal T}_2(\alpha_i)
\nnb \\ &&+
        \frac{1}{qx} (q_\mu x_\nu - q_\nu x_\mu)
        (\varepsilon_\alpha q_\beta - \varepsilon_\beta q_\alpha)
    \int {\cal D} \alpha_i e^{i (\alpha_{\bar q} + v \alpha_g) qx} {\cal T}_3(\alpha_i)
\nnb \\ &&+
        \left. \frac{1}{qx} (q_\alpha x_\beta - q_\beta x_\alpha)
        (\varepsilon_\mu q_\nu - \varepsilon_\nu q_\mu)
    \int {\cal D} \alpha_i e^{i (\alpha_{\bar q} + v \alpha_g) qx} {\cal T}_4(\alpha_i)
                        \right\}, \nnb
\eea
where
$\varphi_\gamma(u)$ is the leading twist 2, $\psi^v(u)$,
$\psi^a(u)$, ${\cal A}$ and ${\cal V}$ are the twist 3 and
$h_\gamma(u)$, $\mathbb{A}$, ${\cal T}_i$ ($i=1,~2,~3,~4$) are the
twist 4 photon DAs, respectively and  $\chi$ is the magnetic susceptibility of the quarks. The measure ${\cal D} \alpha_i$ is defined as
\bea
\int {\cal D} \alpha_i = \int_0^1 d \alpha_{\bar q} \int_0^1 d
\alpha_q \int_0^1 d \alpha_g \delta(1-\alpha_{\bar
q}-\alpha_q-\alpha_g).\nnb
\eea

Explicit form of the photon DAs entering into above matrix elements.

\bea
\varphi_\gamma(u) &=& 6 u \bar u \left( 1 + \varphi_2(\mu)
C_2^{\frac{3}{2}}(u - \bar u) \right),
\nnb \\
\psi^v(u) &=& 3 \left(3 (2 u - 1)^2 -1 \right)+\frac{3}{64} \left(15
w^V_\gamma - 5 w^A_\gamma\right)
                        \left(3 - 30 (2 u - 1)^2 + 35 (2 u -1)^4
                        \right),
\nnb \\
\psi^a(u) &=& \left(1- (2 u -1)^2\right)\left(5 (2 u -1)^2 -1\right)
\frac{5}{2}
    \left(1 + \frac{9}{16} w^V_\gamma - \frac{3}{16} w^A_\gamma
    \right),
\nnb \\
{\cal A}(\alpha_i) &=& 360 \alpha_q \alpha_{\bar q} \alpha_g^2
        \left(1 + w^A_\gamma \frac{1}{2} (7 \alpha_g - 3)\right),
\nnb \\
{\cal V}(\alpha_i) &=& 540 w^V_\gamma (\alpha_q - \alpha_{\bar q})
\alpha_q \alpha_{\bar q}
                \alpha_g^2,
\nnb \\
h_\gamma(u) &=& - 10 \left(1 + 2 \kappa^+\right) C_2^{\frac{1}{2}}(u
- \bar u),
\nnb \\
\mathbb{A}(u) &=& 40 u^2 \bar u^2 \left(3 \kappa - \kappa^+
+1\right) \nnb \\ && +
        8 (\zeta_2^+ - 3 \zeta_2) \left[u \bar u (2 + 13 u \bar u) \right.
\nnb \\ && + \left.
                2 u^3 (10 -15 u + 6 u^2) \ln(u) + 2 \bar u^3 (10 - 15 \bar u + 6 \bar u^2)
        \ln(\bar u) \right],
\nnb \\
{\cal T}_1(\alpha_i) &=& -120 (3 \zeta_2 + \zeta_2^+)(\alpha_{\bar
q} - \alpha_q)
        \alpha_{\bar q} \alpha_q \alpha_g,
\nnb \\
{\cal T}_2(\alpha_i) &=& 30 \alpha_g^2 (\alpha_{\bar q} - \alpha_q)
    \left((\kappa - \kappa^+) + (\zeta_1 - \zeta_1^+)(1 - 2\alpha_g) +
    \zeta_2 (3 - 4 \alpha_g)\right),
\nnb \\
{\cal T}_3(\alpha_i) &=& - 120 (3 \zeta_2 - \zeta_2^+)(\alpha_{\bar
q} -\alpha_q)
        \alpha_{\bar q} \alpha_q \alpha_g,
\nnb \\
{\cal T}_4(\alpha_i) &=& 30 \alpha_g^2 (\alpha_{\bar q} - \alpha_q)
    \left((\kappa + \kappa^+) + (\zeta_1 + \zeta_1^+)(1 - 2\alpha_g) +
    \zeta_2 (3 - 4 \alpha_g)\right),\nnb \\
{\cal S}(\alpha_i) &=& 30\alpha_g^2\{(\kappa +
\kappa^+)(1-\alpha_g)+(\zeta_1 + \zeta_1^+)(1 - \alpha_g)(1 -
2\alpha_g)\nnb \\&+&\zeta_2
[3 (\alpha_{\bar q} - \alpha_q)^2-\alpha_g(1 - \alpha_g)]\},\nnb \\
\tilde {\cal S}(\alpha_i) &=&-30\alpha_g^2\{(\kappa -
\kappa^+)(1-\alpha_g)+(\zeta_1 - \zeta_1^+)(1 - \alpha_g)(1 -
2\alpha_g)\nnb \\&+&\zeta_2 [3 (\alpha_{\bar q} -
\alpha_q)^2-\alpha_g(1 - \alpha_g)]\}.\nnb
\eea
The constants entering  the above DAs are obtained as
\cite{Ball:2002ps} $\varphi_2(1~GeV) = 0$, $w^V_\gamma = 3.8 \pm 1.8$,
$w^A_\gamma = -2.1 \pm 1.0$, $\kappa = 0.2$, $\kappa^+ = 0$,
$\zeta_1 = 0.4$, $\zeta_2 = 0.3$, $\zeta_1^+ = 0$ and $\zeta_2^+ =
0$.


\end{document}